\documentclass[sigconf, authorversion=true]{acmart}

\usepackage{xspace}
\usepackage{threeparttable}
\usepackage{cleveref}
\usepackage{booktabs} 
\usepackage{multirow}

\usepackage{adjustbox}

\usepackage{url}

\usepackage{algorithm}
\usepackage{array}
\usepackage{stfloats}
\usepackage{verbatim}
\usepackage{colortbl}
\usepackage{multicol}

\usepackage{soul} 

\usepackage{comment}

\AtBeginDocument{%
  }

\copyrightyear{2024}
\acmYear{2024}
\setcopyright{acmlicensed}\acmConference[CF '24]{21st ACM International Conference on Computing Frontiers}{May 7--9, 2024}{Ischia, Italy}
\acmBooktitle{21st ACM International Conference on Computing Frontiers (CF '24), May 7--9, 2024, Ischia, Italy}
\acmDOI{10.1145/3649153.3649186}
\acmISBN{979-8-4007-0597-7/24/05}

\begin{document}

\title{A Lightweight Architecture for Real-Time Neuronal-Spike Classification}


\author{Muhammad Ali Siddiqi}
\affiliation{%
  \institution{Lahore University of Management Sciences}
  \country{Pakistan}
}
\email{m.siddiqi@lums.edu.pk}

\author{David Vrijenhoek}
\affiliation{%
  \institution{Delft University of Technology}
  \country{The Netherlands}
}
\email{davidvrijenhoek@gmail.com}

\author{Lennart P. L. Landsmeer}
\affiliation{%
  \institution{Delft University of Technology}
  \country{The Netherlands}
}
\email{l.p.l.landsmeer@tudelft.nl}

\author{Job van der Kleij}
\affiliation{%
  \institution{Delft University of Technology}
  \country{The Netherlands}
}
\email{j.vanderkleij@student.tudelft.nl}

\author{Anteneh Gebregiorgis}
\affiliation{%
  \institution{Delft University of Technology}
  \country{The Netherlands}
}
\email{a.b.gebregiorgis@tudelft.nl}

\author{Vincenzo Romano}
\affiliation{%
  \institution{Erasmus Medical Center, Rotterdam}
  \country{The Netherlands}
}
\email{v.romano@erasmusmc.nl}

\author{Rajendra Bishnoi}
\affiliation{%
  \institution{Delft University of Technology}
  \country{The Netherlands}
}
\email{r.k.bishnoi@tudelft.nl}

\author{Said Hamdioui}
\affiliation{%
  \institution{Delft University of Technology}
  \country{The Netherlands}
}
\email{s.hamdioui@tudelft.nl}

\author{Christos Strydis}
\affiliation{%
  \institution{Erasmus Medical Center, Rotterdam}
  \country{The Netherlands}
}
\email{c.strydis@erasmusmc.nl}

\renewcommand{\shortauthors}{Siddiqi et al.}

\begin{abstract}
Electrophysiological recordings of neural activity in a mouse's brain are very popular among neuroscientists for understanding brain function. One particular area of interest is acquiring recordings from the Purkinje cells in the cerebellum in order to understand brain injuries and the loss of motor functions. However, current setups for such experiments do not allow the mouse to move freely and, thus, do not capture its natural behaviour since they have a \textit{wired} connection between the animal's head stage and an acquisition device. In this work, we propose a lightweight neuronal-spike detection and classification architecture that leverages on the unique characteristics of the Purkinje cells to discard unneeded information from the sparse neural data in real time. This allows the (condensed) data to be easily stored on a removable storage device on the head stage, alleviating the need for wires. Synthesis results reveal a $>$95\% overall classification accuracy while still resulting in a small-form-factor design, which allows for the \textit{free movement} of mice during experiments. Moreover, the power-efficient nature of the design and the usage of STT-RAM (Spin Transfer Torque Magnetic Random Access Memory) as the removable storage allows the head stage to easily operate on a tiny battery for up to approximately 4 days.
\end{abstract}

\ccsdesc[500]{Applied computing~Life and medical sciences}
\ccsdesc[500]{Computer systems organization~Embedded and cyber-physical systems}
\ccsdesc[300]{Hardware~Emerging technologies}

\keywords{Electrophysiological recordings, spike detection, spike classification, Purkinje cells, cerebellum, low-power computing, STT-RAM}


\maketitle

\section{Introduction}
\label{sec:introduction}

\begin{figure}[!t]
    \includegraphics[trim={0cm 0cm 0cm 0cm},clip,scale=0.22]{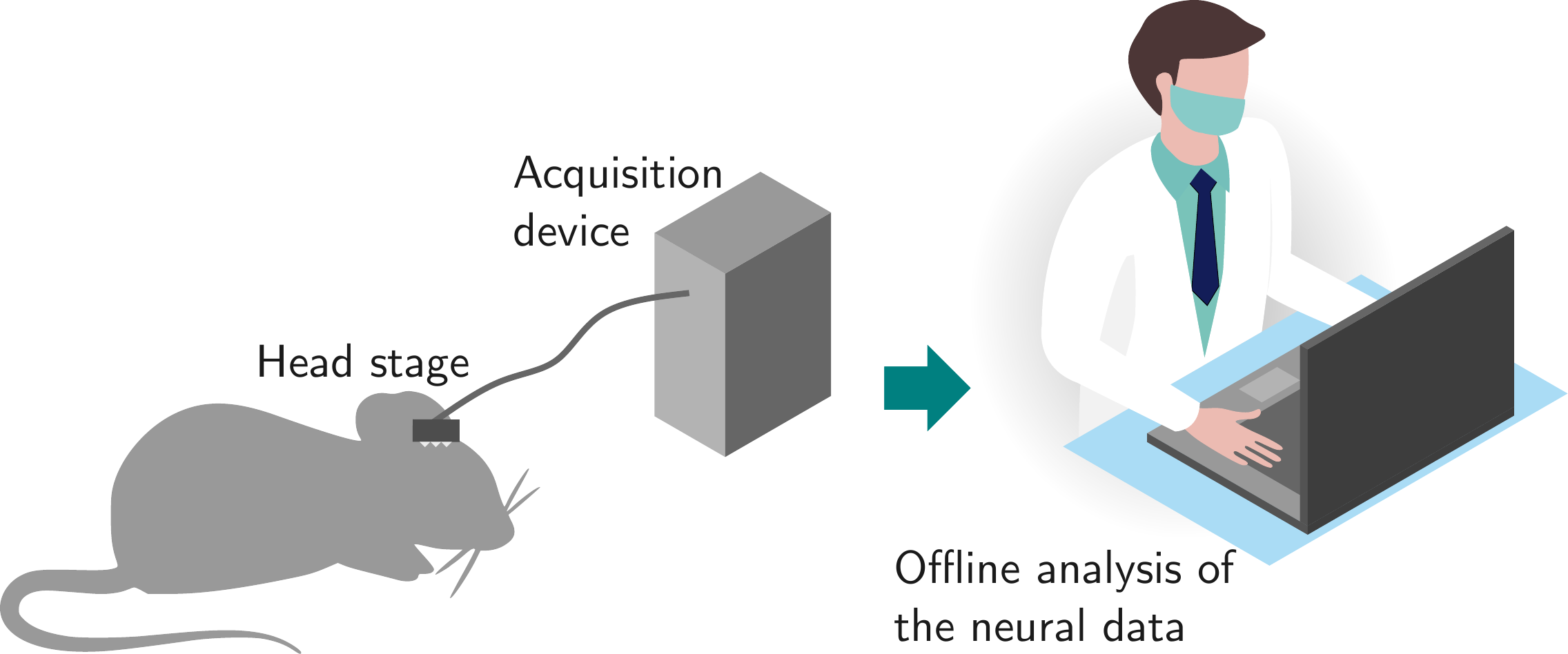}
    \centering
    \caption{A typical wired neural-signal acquisition setup, which limits the free movement of mice.}
    \label{fig:wired-setup}
\end{figure}

The cerebellum is crucial for facilitating motor control and hand-eye coordination, among other critical functionalities~\cite{romano2020functional}. In order to unveil the mechanisms underlying its operation, neuroscientists constantly seek to record and understand its activity in living test subjects. The -- mostly invasive -- nature of these experiments often dictates the use of animal subjects, such as mice~\cite{de2020ninscope}.
One popular methodology relies on electrophysiological recordings of neural activity in various types of cells, especially the Purkinje cells in the cerebellum because of their critical role in motor coordination~\cite{romano2020functional}.
In current setups, a \textit{wired} connection is used between the mouse and an acquisition device, as shown in Figure~\ref{fig:wired-setup}.
However, such setups do not mimic natural conditions since the wires do not allow \textit{free movement} of mice.
Therefore, in essence, we need to get rid of these wires to enable more \textit{realistic} neuroscientific experiments.

Several \textit{wireless} head stages for different mammals have been proposed~\cite{freelynx2021manual,bilodeau2021wireless,cereplex2020manual,ecube2020,gagnon2016wireless,grand2013long}.
However, they have \textit{at least} one of these two limitations: (1) They are too heavy and, thus, are only suitable for larger animals (in the case of~\cite{freelynx2021manual,bilodeau2021wireless,cereplex2020manual,grand2013long}).
Simply put, the whole head stage, including the battery, needs to be $<$ 3 grams for mice for the aforementioned cerebellum experiments \cite{de2020ninscope}.
(2) The head stages can record for only a short period of time, e.g., 30 and 105 minutes in the case of~\cite{ecube2020} and~\cite{gagnon2016wireless}, respectively, whereas the aforementioned experiments require up to 24 hours of neural recordings.
This is because the study of learned motor control, information processing, memory consolidation, interactions among distributed brain regions etc., requires long timescales~\cite{chung2019high}, see Table~\ref{tab:duration}.
In short, \textit{suitable} wireless head stages for \textit{mice} do not yet exist, which is primarily due to the difficulty in constructing on-site neural-data processing and a wireless transceiver in a small form factor.
For instance, algorithms for processing neural data (including that from Purkinje cells), such as~\cite{sedaghat2021p,markanday2020using,zur2019using,luan2018compact}, are generally designed in software for offline analysis and have not yet reached the required form factor to be incorporated within a mouse's head stage.

This work addresses both of the aforementioned limitations by introducing a novel approach for real-time detection and classification of neuronal spikes in the sampled neural data from Purkinje cells.
The purpose of this scheme is to reduce the dimensionality of this sparse data so that only a small but essential part is stored on a removable storage within the head stage, which alleviates the need for wires.
This heavily condensed stored data can then be retrieved after an experiment for offline analysis. In this way, our scheme allows to reach the ultimate \textit{goal} of conducting \textit{long-duration} experiments involving \textit{freely moving} mice.
In essence, this work makes the following key contributions:

\begin{itemize}
  \item A \textit{neuronal-spike classification scheme} that throws out unnecessary information from the stream of data from the Purkinje cells, taking advantage of their unique characteristics, which simplifies the data-storage requirements.
  \item A \textit{lightweight system architecture} that consists of a controller that orchestrates the data flow from the spike detector to the classification module and ultimately stores the classified data in a non-volatile memory (STT-RAM).
  \item A \textit{small-form-factor synthesized CMOS design} that enables low-power operation for a prolonged duration of time while staying within the head-stage size requirements.
\end{itemize}

The rest of the paper is organized as follows. Section~\ref{sec:background} provides a brief neuroscientific background of the experiments involving cerebellar recordings. Section~\ref{sec:proposed-solution} explains our proposed scheme followed by the results in Section~\ref{sec:results}. We draw overall conclusions in Section~\ref{sec:conclusions}.

\section{Background}
\label{sec:background}

\subsection{Neuroscientific Background}

Techniques developed during the second half of the nineteenth century allowed neuroscientists to investigate the electrical activity of an animal brain directly using electrophysiological recordings.
A particular area of interest is the use of such techniques to acquire and analyze the neuronal activity of the Purkinje cells in the cerebellum.
A detailed list of animal behaviours that could be studied using such experiments is provided in Table~\ref{tab:duration}.
These experiments shed light on how the activity of a particular brain area, in this case the cerebellum, relates (and possibly give rise) to a particular sensory-motor or cognitive function.
This could lay the foundation for a better understanding of the brain and could, potentially, have application in treating brain injuries and loss of motor functions~\cite{romano2020functional}. 
Mice are usually preferred for these studies since they are easy to manage and very well-suited for genetic manipulations, which significantly increases the number of research questions that can be addressed.

Current experiments involving neural recordings from mice have one major constraint: the animals are head-fixed within the experimental setup that only allows them some body movement, which is unnatural. For instance, when a certain neuronal signal appears to be associated with a particular movement, it is not possible to establish if that neuronal signal is also associated with the movement of other parts of the body that are restrained or immobilized.
The size and weight of the head stage (i.e., it should be roughly $<$3 grams for mice) and the recording duration (i.e., it should support 24-hour-long sessions) constitute the primary challenges in developing technologies that meet the goal of ensuring the natural states of animals during experiments. Significant effort has been devoted over the past decade to addressing these challenges, aiming to enable freely moving animal recordings, as discussed next.

\begin{table*}[!t]
    \centering
    \caption{Examples of behaviours and paradigms that could be studied using neuronal recordings involving freely moving mice.}
    \label{tab:duration}
    \begin{adjustbox}{max width=17.5cm}
    \begin{threeparttable}
    \begin{tabular}{llll}
        \toprule
        Behaviour & Main topics to be studied & Duration & Examples \\
        \midrule
        Rhythmic movements & Motor control & 5-10 minutes & Respiration, whisking, locomotion etc. \\
        Compensatory eye movement & Simple reflex & 30-60 min & VOR and OKR \\
        Social interaction & Cognitive functions & 30-60 min & Two chambers test \\
        Sleep & Coherent oscillation & 12-24 hours & Recording of spontaneous sleep \\
        Grasping task & Learned motor control & Multiple sessions across 2-3 days$^*$ & Food pellet or water reaching task \\
        Adaptation of compensatory eye movement & Memory formation & Multiple sessions across 4-5 days$^*$ & VOR/OKR adaptation \\
        Eyelid conditioning & Associative learning & Multiple sessions across 6-7 days$^*$ & Delay eyelid conditioning \\
        Sensory discrimination task & Cognitive learning & Multiple sessions across 16-20 days$^*$ & Directional licking task \\
        \bottomrule
    \end{tabular}
	\begin{tablenotes}[flushleft]
		\item `VOR': Vestibulo-ocular Reflex, `OKR': Optokinetic Response
		\item $^{*}$A recording session can take up to 24 hours.
	\end{tablenotes}
    \end{threeparttable}
    \end{adjustbox}
\end{table*}

\subsection{Related Work}
\label{sec:related-work}

Bilodeau et al.~\cite{bilodeau2021wireless} and Gagnon-Turcotte et al.~\cite{gagnon2016wireless} present head stages that utilize Spartan-6 FPGAs for real-time processing of recorded neural signals and wireless transceivers to transmit the reduced data to an external device (base station). These head stages are suitable for mice due to their reasonable weight. However, their maximum recording durations, less than 1.75 hours, are too short for the cerebellar experiments highlighted in Table~\ref{tab:duration}.
Grand et al.~\cite{grand2013long} propose a head stage design where recorded neural data is directly transmitted wirelessly to an external device without any onboard processing. Their system allows for an incredible 72-hour recording duration. However, the large size and weight of the complete system make it suitable only for large animals, such as monkeys.
Luan et al.~\cite{luan2018compact} also employ FPGA-based dimensionality reduction, similar to~\cite{bilodeau2021wireless, gagnon2016wireless}. Their design allows for an impressive 24-hour recording duration. However, like~\cite{luan2018compact}, it is only suitable for large animals, in their case, cats.

When focusing on the dimensionality reduction of neural data from Purkinje cells, there is a limited body of work available. Notably, all these works are applicable to \textit{offline} software implementations, wherein they process raw digitized data retrieved from a wired setup.
However, since these works are intended to be executed on larger computing devices (e.g., PCs, GPUs, etc.), they, in their current form, cannot be executed on a small head stage due to the dearth of computing resources available onboard.
These works encompass the research by Sedaghat-Nejad et al.~\cite{sedaghat2021p}, Markanday et al.~\cite{markanday2020using}, and Zur et al.~\cite{zur2019using}.

In addition to literature, commercial solutions are also available. For example, the FreeLynx wireless acquisition system from NeuraLynx~\cite{freelynx2021manual} provides a convenient setup for enabling freely-moving animal recordings. However, its smallest head-stage configuration has a weight of 21 g, making it suitable only for larger animals, such as rats and monkeys. A similar issue is observed with the CerePlex Exilix system from Blackrock Microsystems~\cite{cereplex2020manual}, which features a 9.87 g head stage and a limited recording duration of up to 2.5 hours.
The eCube wireless head stage from White Matter~\cite{ecube2020} shows promise with a 2.5 g head stage, including the battery, making it suitable for mice. Unfortunately, the recording duration is only 30 minutes, which is inadequate for the aforementioned cerebellar experiments. Hence, there is still a pressing need for new head-stage technologies that are small, light, low-power, and wireless.

\section{Proposed Solution}
\label{sec:proposed-solution}

\begin{figure*}[!t]
    \includegraphics[trim={0cm 0cm 0cm 0cm},clip,scale=0.24]{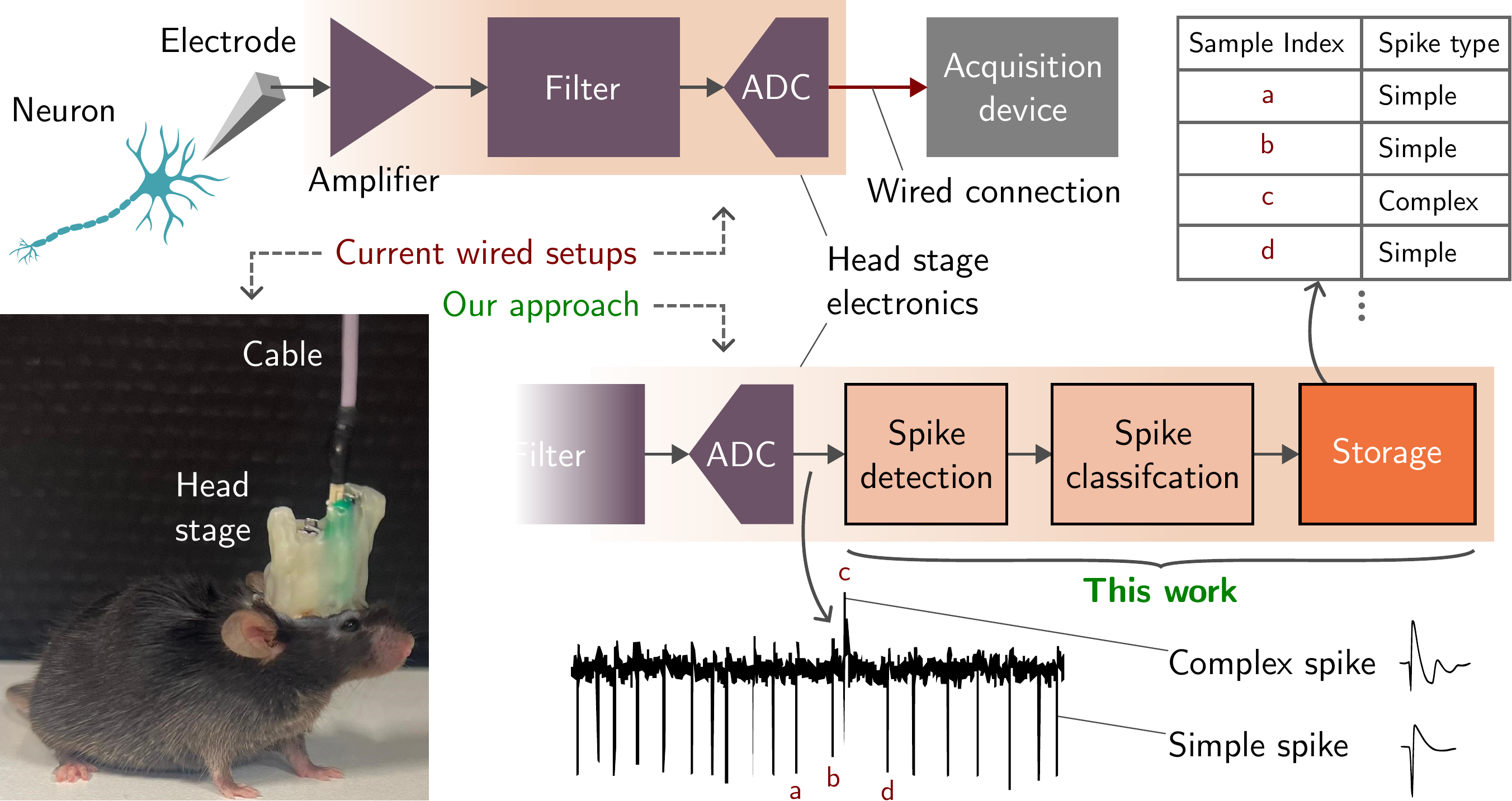}
    \centering
    \caption{Current wired setups vs. our approach, which gets rid of the wire and enables free movement of mice.}
    \label{fig:existing-vs-new-approach}
\end{figure*}

\begin{figure*}[!t]
    \includegraphics[trim={0cm 0cm 0cm 0cm},clip,scale=0.85]{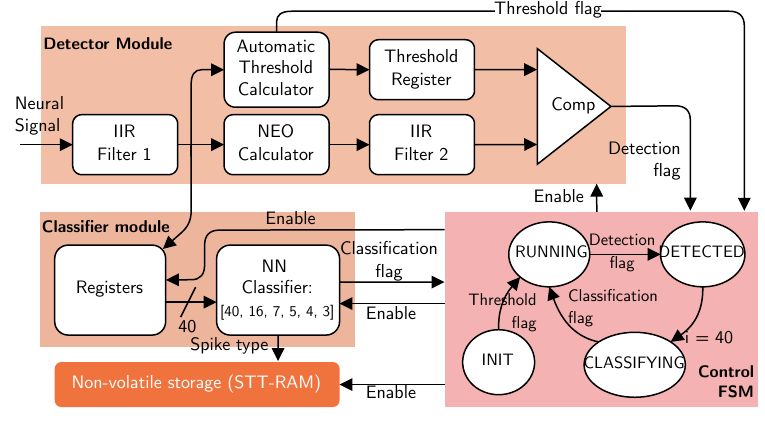}
    \centering
    \caption{Proposed system architecture.}
    \label{fig:architecture}
\end{figure*}

Our approach deviates from using a wireless transceiver to get rid of the wires during neural-signal acquisition from mice. This is because a wireless transceiver involves a complex design that impacts system reliability and, of course, power consumption.
Instead, we take advantage of the fact that the signal analysis related to the aforementioned cerebellar experiments is usually done \textit{after} the recording phase, i.e., offline. As a result, our proposal to eliminate the wires is to store the neural data on a removable storage on the head stage, which can then be retrieved for offline analysis (see Figure~\ref{fig:existing-vs-new-approach}).
However, this approach, as is, does not scale well for larger number of channels and longer experiments (i.e., up to 24 hours) since it would require a significantly large amount of storage.

We solve this problem by leveraging the unique characteristics of the Purkinje cells and keeping only the information of interest from the very sparse neural recording.
More specifically, this crucial information (for the experiments discussed in Section~\ref{sec:introduction}) is (a) the time when a spike occurred and, (b) the type of spike, i.e., \textit{simple} or \textit{complex} (since a Purkinje-cell spike can be of one of these two types~\cite{markanday2020using}, as shown in Figure~\ref{fig:existing-vs-new-approach}).
This results in storing only a condensed version of the complete recording (see Figure~\ref{fig:existing-vs-new-approach}, top right), which significantly eases up storage requirements and enables a small-form-factor implementation.

An overview of the proposed scheme is shown in Figure~\ref{fig:architecture}. The detector module, which is based on~\cite{yang2016hardware}, detects the occurrence of a spike in an input stream of digitized neural data. After this detection, the classifier module is enabled, which determines whether the spike is simple or complex. Both these modules are explained next.

\subsection{Spike Detection}
\label{sec:spike-detection}

The spike-detection module is based on the work of Yang et al.~\cite{yang2016hardware}. This specific design is chosen because it can support the Purkinje-cell firing rate of around 100 Hz while having a small form factor and being extremely power efficient.

The digitized input samples are first filtered using a 1st-order exponential IIR filter, which smooths the input signal to mitigate high-frequency noise components that adversely effect the performance of the spike detector. The filtered signal is then fed to a non-linear energy operator (NEO)~\cite{mukhopadhyay1998new} and an accompanying 1st-order exponential IIR filter to get the instantaneous estimation of signal energy, which is then compared with a calculated threshold value. If the energy estimate (NEO value) is higher than the threshold then a spike is detected, and vice versa.
The threshold calculator is able to dynamically (re)calculate the threshold and minimize the performance loss in case of changes to the input signal, e.g., due to the worsening of the noise level.

\subsection{Spike Classification}
\label{sec:spike-classification}

After the detection of a spike, the spike waveform (i.e., a certain number of samples after the detection event) is saved in order for it to be classified as either being simple or complex.
The length of this waveform is set as 40 samples to sufficiently capture the whole spike duration ($\sim$1.7 ms~\cite{markanday2020using}) considering the sampling frequency of 24.414 kHz.
We chose to employ a neural-network (NN)-based design for this classifier. 
This is because NNs tend to generalize well when there is plenty of training data, as in our case.
Such a classifier would, therefore, be robust enough to handle varying signal conditions. For instance, the recorded data set used in this work exhibits significant variations in terms of noise floor, recording-electrode drift, saturated sample points, and data offsets.

\subsubsection{Choice of NN Topology}
\label{sec:nn-topology}

The highly resource-constrained nature of the head stage necessitates the use of a lightweight NN.
Thankfully, the classification problem itself is a \textit{ternary} classification problem (i.e., classifying a spike as simple, complex or neither), which means that the NN needs only three output neurons.
We chose to tackle this relatively simple classification problem using a \textit{multilayer perceptron}, i.e., a fully connected feedforward artificial neural network.

Keeping the NN lightweight also implies finding an optimal trade-off between the classification accuracy and area/energy consumption.
Thus, an extensive design space exploration (DSE), which will be discussed in Section~\ref{sec:dse}, was conducted concerning the choice of NN topology, including the number of hidden layers and neurons per layer, and its impact on classification accuracy.
The NN has 40 inputs to correspond with the aforementioned 40-sample spike waveform.

\subsubsection{Choice of Activation Functions}
\label{sec:activation-function}

For the hidden layers, a multitude of activation functions exist to allow the network to learn the necessary nonlinearities in the classification problem.
Popular choices include various sigmoidal-shaped functions and ReLU (rectified linear unit). 
As hardware implementation of nonlinear functions is very costly, it was decided to choose ReLU for the hidden layers since it is the least computationally-expensive option.

To calculate the final probabilities in the \textit{output} layer of an NN for multi-class classification problems like this spike-classification task, the Softmax function is commonly employed.
However, this is a very costly operation in hardware, which would also require IEEE floating point support.
Instead, Softmax can be moved to the loss function, effectively making the network output log probabilities (logits) instead of just probabilities. 
This can be achieved using a simple linear activation
function that caries no computational cost during inference.
Since the above natural logarithm is a monotonically increasing function of probability, the output class is determined by selecting the index of the neuron with the highest value, just like in the case of using normal probabilities.

\subsubsection{NN Quantization}
\label{sec:quanitzation}

The NN was initially designed and tested in software before implementing it in hardware (see Section~\ref{sec:setup} for more details on the tool flow). Conventionally, software NN implementations use floating-point arithmetic which adds to the computational burden of the classifier. Through the quantization technique presented in~\cite{jacob2018quantization}, we transformed the NN to integer format to increase the efficiency of the inference. Following this technique, the weights and activation values of the neurons were mapped to signed 8-bit values.

\subsection{Control}
\label{sec:control}

Given the fact that the occurrence of Purkinje-cell spikes is relatively rare (i.e., roughly 100 Hz on average) compared to the sampling frequency, constantly running all of the modules in the proposed architecture (Figure~\ref{fig:architecture}) is unnecessary.
In fact, once the threshold has been calculated for the detection, the only block in the system that runs continuously is the NEO calculator and the comparator. The classifier is only enabled once a spike has been detected.
Furthermore, there is no need to detect another spike during the classification execution on the initial spike as there is a sufficient time gap ($>$2 ms) between two consecutive spikes.
To control the information flow and enabling/disabling modules, a finite-state machine (FSM) was designed, whose state diagram is shown in Figure~\ref{fig:architecture}.

The system is in the \textit{INIT} state when it is initially started or is reset. During \textit{INIT} a threshold is calculated for the spike detection and all the other subsystems are disabled. When the threshold calculator converges to a threshold, a flag is set. At this point, the FSM transitions to the \textit{RUNNING} state in which the NEO calculator is enabled and spike detection is performed. At this point, there is no need to enable the threshold calculator or classifier. When a calculated NEO value exceeds the threshold, a detection flag is set and the FSM transitions to the \textit{DETECTED} state. During this state, the next 40 filtered input samples are saved. Here, the precision of the samples is reduced to 8 bits as mentioned earlier in Section~\ref{sec:quanitzation}. After these 40 cycles, the FSM transitions to the \textit{CLASSIFYING} state in which the NN inference is performed and the classified result is stored in the storage. Upon completion, the system transitions back to the \textit{RUNNING} state.

\subsection{Storage}
\label{sec:storage}

One of the important aspects of our approach is the use of on-board storage to store the reduced-size (i.e., classified) data. 
Since this memory needs to be removed after an experiment to retrieve the recordings, it needs to be non-volatile.
Another reason for employing a non-volatile memory is to allow it to stay powered down when the head stage is not classifying, which cuts out its leakage power and improves battery life.
For the proposed scheme, we chose the \textit{Spin Transfer Torque Magnetic Random Access Memory} (STT-RAM). This is because apart from non-volatility, STT-RAM has a small form factor, low access latency and energy, high endurance, CMOS compatibility, high maturity and immunity to soft-errors due to radiations~\cite{bertolazzi2019nonvolatile}. 

Table \ref{tablmemories} illustrates a comprehensive comparison between STT-RAM and various memory technologies, encompassing both conventional and emerging non-volatile memories. Conventional memories like SRAM and DRAM are volatile, necessitating a continuous power supply. SRAM notably faces leakage issues, and DRAM necessitates periodic refresh cycles, rendering it energy-inefficient. Flash memory, another conventional technology, operates at relatively high voltages. Among the emerging non-volatile technologies, RRAM encounters endurance problems, while PCM demands high voltage for switching. Although STT-RAM boasts numerous advantages, it has the drawback of requiring a constant current for writing, resulting in slightly higher energy consumption and latency compared to the traditional SRAM technology. However, this drawback is offset by SRAM's leakage issues. Due to all these benefits, STT-RAM stands out as the only commercially available emerging non-volatile technology in the market due to its overall performance and reliability \cite{Everspin, nair2018defect}. 
Furthermore, since we do not need STT-RAM's prolonged data-retention (of up to 10 years) for the targeted experiments, we take advantage of the tunability of its thermal-stability factor by which we shorten the retention time in return for further improving its energy efficiency and access latency~\cite{smullen2011relaxing}. 
These features make STT-RAM very suitable for our experiments.

\begin{table*}[!t]
        \centering
	\caption{Comparison of bit-cell design metrics for various memory technologies (data obtained from~\cite{salahuddin2018era,oboril2015evaluaion}).}
	\label{tablmemories}
	\begin{tabular}{lllllll}
		\hline
            Metrics & SRAM & DRAM  & Flash & RRAM & STT-RAM & PCM \\   
 
		\hline
		Size ($F^2$) &  \cellcolor{red!25} 120--150 & \cellcolor{green!25} 10--30 & \cellcolor{green!25} 10--30 & \cellcolor{green!25} 10--30 & \cellcolor{green!25} 10--30 & \cellcolor{green!25} 10--30\\ 
		Volatility & \cellcolor{red!25} Yes & \cellcolor{red!25} Yes & \cellcolor{green!25} No & \cellcolor{green!25} No & \cellcolor{green!25} No & \cellcolor{green!25} No\\ 
		Write energy & \cellcolor{green!25} $\sim$fJ & \cellcolor{green!25} $\sim$10 fJ & \cellcolor{red!25} $\sim$100 pJ & \cellcolor{yellow!25} $\sim$1 pJ & \cellcolor{yellow!25} $\sim$1 pJ & \cellcolor{red!25} $\sim$10 pJ\\ 
		Write speed & \cellcolor{green!25} $\sim$1 ns & \cellcolor{yellow!25} $\sim$10 ns & \cellcolor{red!25} 0.1--1 ms & \cellcolor{yellow!25} $\sim$10 ns & \cellcolor{yellow!25} $\sim$5 ns & \cellcolor{yellow!25} $\sim$10 ns\\ 
		Read speed & \cellcolor{green!25} $\sim$1 ns & \cellcolor{green!25} $\sim$3 ns & \cellcolor{red!25} $\sim$100 ns & \cellcolor{yellow!25} $\sim$10 ns & \cellcolor{yellow!25} $\sim$5 ns & \cellcolor{yellow!25} $\sim$10 ns\\ 
		Endurance & \cellcolor{green!25} $10^{16}$ & \cellcolor{green!25} $10^{16}$ & \cellcolor{red!25} $10^4$--$10^6$ & \cellcolor{red!25} $10^7$ & \cellcolor{green!25} $10^{15}$ & \cellcolor{yellow!25} $10^{12}$\\
		Scalability & \cellcolor{yellow!25} medium & \cellcolor{yellow!25} medium & \cellcolor{yellow!25} medium & \cellcolor{green!25} high & \cellcolor{green!25} high & \cellcolor{green!25} high\\ 
		\hline
	\end{tabular}
\end{table*}

\subsection{Post-processing}
\label{sec:post-processing}

\begin{figure}[!t]
    \includegraphics[trim={0cm 0cm 0cm 0cm},clip,scale=0.25]{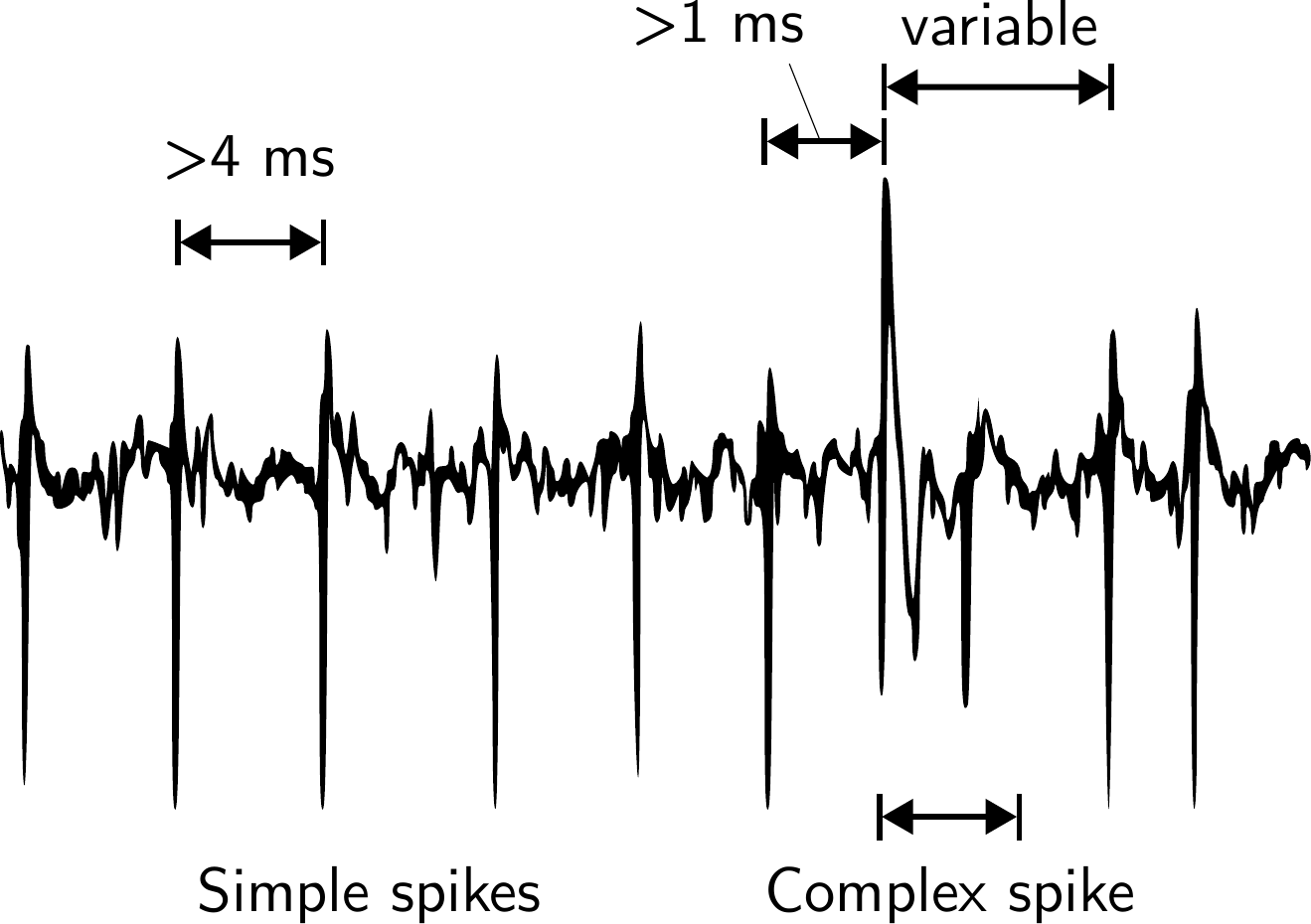}
    \centering
    \caption{Relative timing information between complex and simple spikes, which can aid in the offline post-processing}
    \label{fig:deadlines}
\end{figure}

The calculated NEO value of the spike detector can sometimes exceed the threshold several times in a short duration, leading to one actual spike causing multiple spike detections. To address this limitation of the spike-detection algorithm, a post-processing step is carried out \textit{offline}, i.e., on the reduced data retrieved from the head-stage storage. This step leverages the relative spike timing between the spikes, as illustrated in Figure~\ref{fig:deadlines}. The time interval between these (false) detection events often falls within the sub-millisecond domain, which is anatomically implausible for Purkinje cells: Neurons require time to accumulate the necessary charge (i.e., ion concentration) to generate a spike, typically exceeding 4 ms~\cite{de2011spatiotemporal}. 
As a result, a dead zone is introduced in post-processing after a detection event. However, since the minimum spike interval pertaining specifically to simple spikes remains relatively consistent, the dead zone is applied \textit{only} after the classification of simple spikes, and any detection events within this period are subsequently discarded.

\section{Results}
\label{sec:results}

\subsection{Experimental Setup}
\label{sec:setup}

\begin{figure}[!t]
    \includegraphics[trim={0cm 0cm 0cm 0cm},clip,scale=0.64]{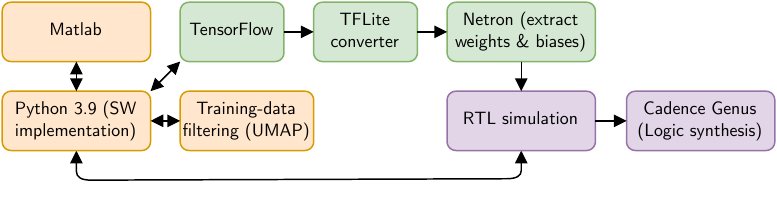}
    \centering
    \caption{Employed tool flow}
    \label{fig:toolflow}
\end{figure}

An overview of the employed tool chain is shown in Figure~\ref{fig:toolflow}.
The data used to design, train and verify the system was formatted in Matlab.
This data set was acquired using a wired setup (see Figure~\ref{fig:existing-vs-new-approach}) involving the Mini-Amp-64 head stage from Cambridge NeuroTech~\cite{cambridge2020manual} with the ADC sampling frequency of 24.414 kHz and a resolution of 10 bits.
To enable flexibility and multiple DSE iterations, the spike detector and classifier were first implemented in software (Python).
Python was employed for this purpose because of its ease of use and well-supported libraries (especially in regards to the NN-based classifier). 
The hardware was described in VHDL and simulated using Xilinx Vivado. The classifier NN training was done using TensorFlow. The training data itself was sanitized using Uniform Manifold Approximation and Projection for dimension reduction (UMAP)~\cite{mcinnes2018umap}.
Subsequently, TFLite was used to optimize (quantize) the TensorFlow models~\cite{jacob2018quantization}. The Netron software~\cite{netron2022} was used to visualize the neural network and extract the weights and biases, which were then used to create a hardware description of the classifier. For our evaluation, we have synthesized the design on Cadence Genus, using the 45-nm NanGate Open-Cell library~\cite{nangate2019}.
The design metrics for the STT-RAM were extracted using the NVSim tool~\cite{dong2012nvsim}.

\subsection{Spike-Classifier-NN Training}
\label{sec:training}

The acquired neural-recording contains around $10^6$ spikes, which were \textit{manually} annotated offline in order to create a reference for training the spike-classifier NN.
40 samples after every NEO-filter-detected sample were saved in order to cover the rough spike duration (as mentioned in Section~\ref{sec:spike-classification}).
Based on the annotated data, these detected waveforms were labeled as \textit{Complex Spike (CS)}, \textit{Simple Spike (SS)}, or \textit{False positive (F)} (for detected signals that did not correspond to annotated spikes).
Since these samples were in a 10-bit format (ADC resolution), the whole waveform was divided by 4 since the NN requires 8-bit inputs.

Furthermore, since the occurrence of complex spikes is far less frequent than that of simple spikes (i.e., there is a large class imbalance), the training data set was re-sampled to obtain a balanced set of simple and complex spikes, and false positives (5953 SS, 5953 CS and 5948 F samples, respectively). This process prevents the NN from being biased towards simple spikes.
This reduced data set was further split randomly into the \textit{training set} (80\%) and \textit{test set} (20\%).
The training was performed over 100 epochs and \textit{early stopping} with a 90\%/10\% training/validation split on each fold.

\begin{figure}[!t]
    \includegraphics[trim={0cm 0cm 0cm 0cm},clip,scale=0.315]{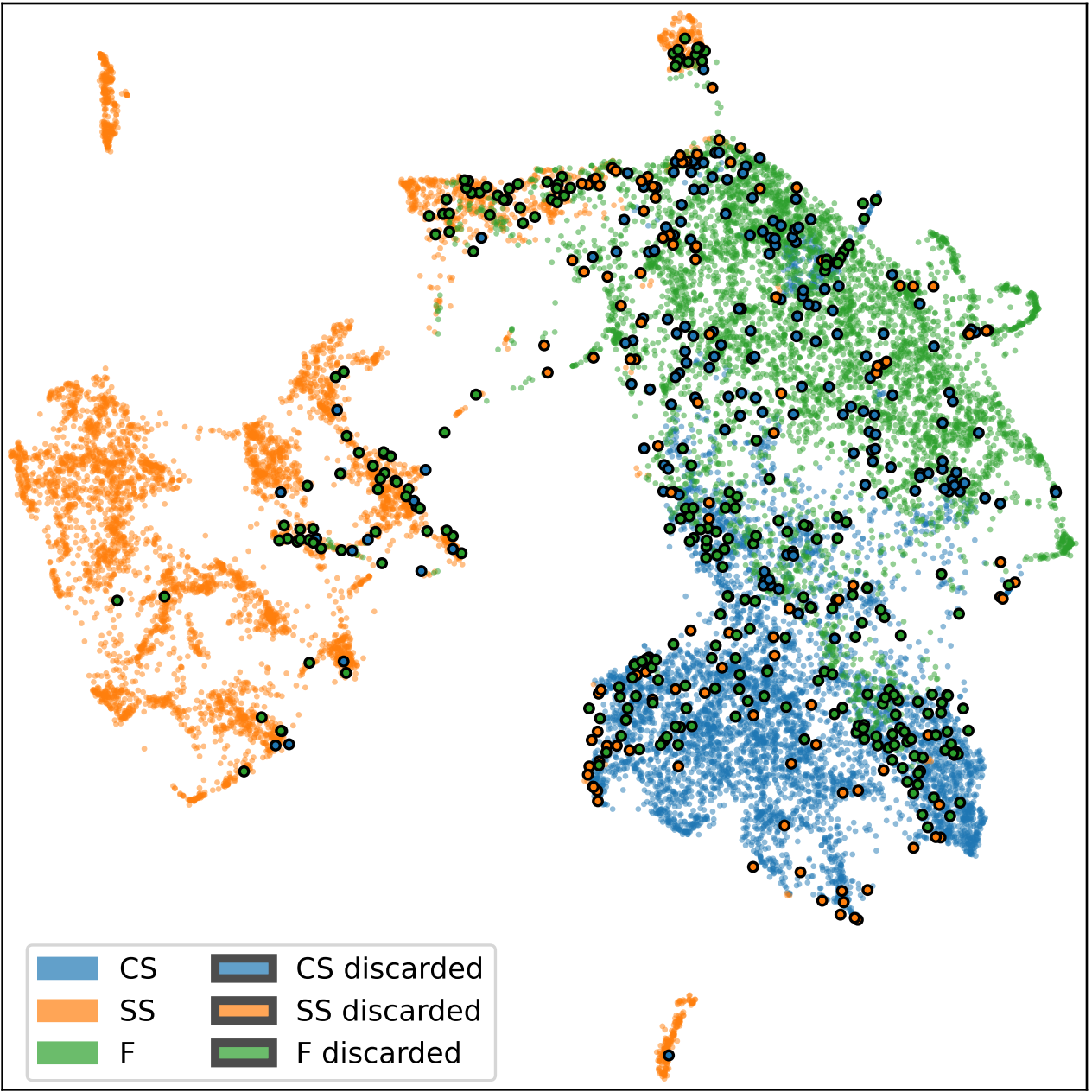}
    \centering
    \caption{UMAP projection of the \textit{class-balanced} dataset and highlighted outliers to be discarded during training}
    \label{fig:umap}
\end{figure}

The training samples underwent another round of filtering by comparing them against the output of UMAP, which was performed on the aforementioned \textit{balanced} data set (in two dimensions). It was found that a small amount of samples seemed to be outliers or were very similar to other classes, which could be due to possible mislabeling or missing spikes altogether during manual annotation.
To avoid confusing the training protocol with these outliers, we discarded samples when the 10 nearest neighbours in the UMAP space contained 9 or more differently-labeled samples at the time of training (Figure~\ref{fig:umap}).
It is important to note that the outliers were not discarded during the \textit{testing} of the NN (Section~\ref{sec:testing}).

The loss function used for training is the \textit{multi-class cross-entropy}, calculated directly from logits to remove Softmax-computation overhead from the network output layer. As the name suggests, this loss function is well-suited for a ternary-classification problem, such as this one. 
The weights were updated using the popular Adam optimizer~\cite{kingma2014adam}.

\subsection{Classifier Design Space Exploration}
\label{sec:dse}

The DSE of the spike-classifier NN was performed to find an optimal trade-off between the classification accuracy and area/energy consumption, which in turn depends on the NN topology (i.e., the number of hidden layers and the number of neurons per layer).
To gauge the classification accuracy, we employed the classical formula of $\frac{T_n + T_p}{T_n+T_p+F_n+F_p}$, where $T_n$ and $T_p$ are the true negatives and positives, respectively and $F_n$ and $F_p$ are the false negatives and positives, respectively.
To obtain the most ideal NN topology, a 10-fold cross-validated grid search was performed over NN architecture and regularization. 
The search included all possible hidden layer configurations between 0 and 4 layers, in descending neuron counts for each layer, i.e., 1--40, 1--20, 1--10, and 1--10, respectively.
Furthermore, the grid search contained a 0.01 or 0.001 regularization constraint for the weight-matrix orthogonality.
It should be noted that all accuracy estimates were performed \textit{after quantization}, i.e., it was incorporated into the DSE.

As CS classification is the hardest task, the most optimal network after the grid search was selected by first choosing those with a CS class accuracy greater than 90\% within a 95\% confidence interval. This criterion was applied to closely match the performance of the recent software-based technique~\cite{markanday2020using} discussed in Section~\ref{sec:introduction}. Subsequently, the network with the lowest computational complexity was selected, as estimated by $\sum_{i\in \text{layers}} n(i)n(i+1)$, where $n(i)$ represents the size of the layer.
The network-architecture optimization results are shown Table~\ref{tab:dse}.
The optimal NN topology consisted of four hidden layers with 16, 7, 5, and 4 neurons, respectively, along with the orthogonal weight regularization set to 0.01.

\begin{table}[!t]
    \centering
    \caption{Network-architecture search results}
    \label{tab:dse}
    \begin{adjustbox}{max width=8.2cm}
    \begin{threeparttable}
    \begin{tabular}{llllll}
        \toprule
        Network  & RF & CS$^{*}$ & SS$^{*}$ & F$^{*}$ & Complexity$^{**}$ \\
        Architecture &  &  & &  & \\
        \midrule
        40, 2, 3 & 0.001 & 85.4\% & 93.4\% & 83.0\% & 86 \\
        40, 2, 3 & 0.010 & 86.4\% & 93.7\% & 82.2\% & 86 \\
        40, 4, 3, 3 & 0.010 & 88.0\% & 94.3\% & 84.4\% & 181 \\
        40, 5, 5, 2, 3 & 0.010 & 88.9\% & 93.5\% & 82.8\% & 241 \\
        40, 7, 7, 4, 3, 3 & 0.010 & 90.0\% & 93.5\% & 83.0\% & 378 \\
        40, 8, 8, 3, 3, 3 & 0.001 & 90.6\% & 94.2\% & 84.0\% & 426 \\
        40, 14, 10, 4, 3, 3 & 0.010 & 91.3\% & 94.8\% & 86.2\% & 761 \\
        40, 16, 7, 5, 4, 3 & 0.010 & 91.7\% & 94.5\% & 85.1\% & 819 \\
        40, 28, 14, 8, 6, 3 & 0.010 & 93.0\% & 94.5\% & 89.0\% & 1690 \\
        \bottomrule
    \end{tabular}
        \begin{tablenotes}[flushleft]
		  \item `RF': Regularization Factor
            \item[$^{*}$] Mean post-quantization accuracy calculated across 10-fold cross-validation
            \item[$^{**}$] Complexity defined as $\sum_{i\in \text{layers}} n(i)n(i+1)$.
        \end{tablenotes}
    \end{threeparttable}
    \end{adjustbox}
\end{table}

\subsection{Classifier Testing}
\label{sec:testing}

\begin{figure}[!t]
    \includegraphics[trim={0cm 0cm 0cm 0cm},clip,scale=.34]{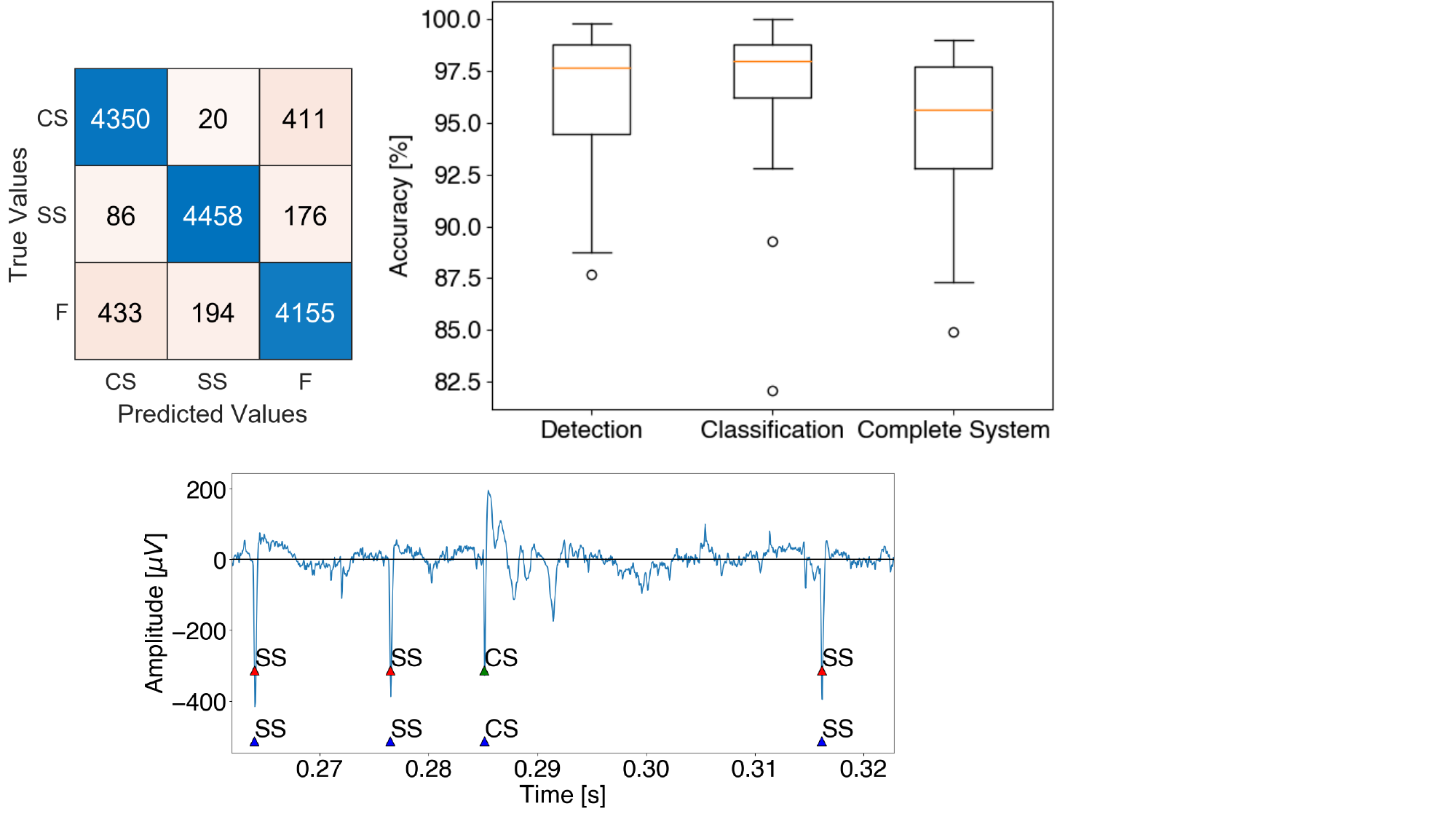}
    \centering
    \caption{Top left: Final confusion matrix of the classifier NN on the test data set. Top right: Accuracy of our scheme for different neural recordings. Bottom: Manual classification vs. automated classification using our scheme. Blue triangles denote the manually-labeled spikes and the red/green triangles denote the detected simple (SS) and complex (CS) spikes, respectively.}
    \label{fig:confusion-accuracy}
\end{figure}

Finally, the aforementioned optimal network was retrained and quantized on the full training set, and was subsequently tested on the complete \textit{test data set} (which remains unseen until now), leading to the confusion matrix shown in Figure~\ref{fig:confusion-accuracy} (top left).
We can see that both the CS (93.35\%) and SS (96.67\%) classification accuracies satisfy the target of $>$90\%.

Figure~\ref{fig:confusion-accuracy} (top right) shows that for the selected NN topology the median accuracy values of the individual blocks and the complete system are greater than 95\%.
Figure~\ref{fig:confusion-accuracy} (bottom) shows a snippet of a neural recording in which our scheme detects and classifies the spikes with the same outcome as that of the manual annotation.

\subsection{Energy- and Area-Efficiency Analysis}

\begin{table}[!t]
    \centering
    \caption{Implementation results}
    \label{tab:implementation-numbers}
    \begin{tabular}{lccc}
        \hline
          & Detector & Classifier &  Storage (STT-RAM) \\ 
        \hline
        Energy$^{*}$ (nJ) & 4.46 & 311 & 0.28  \\
        Area (mm$^\textrm{2}$) & 0.006 & 0.081 & 25.91$^{**}$  \\
        \hline
    \end{tabular}
	\begin{tablenotes}
		\footnotesize
		\item $^{*}$One complete detection-classification-storage cycle
		\item $^{**}$For 32-MB capacity to support 24-hour experiments
	\end{tablenotes}
\end{table}

\begin{figure}[!t]
    \includegraphics[trim={0cm 0cm 0cm 0cm},clip,scale=0.30]{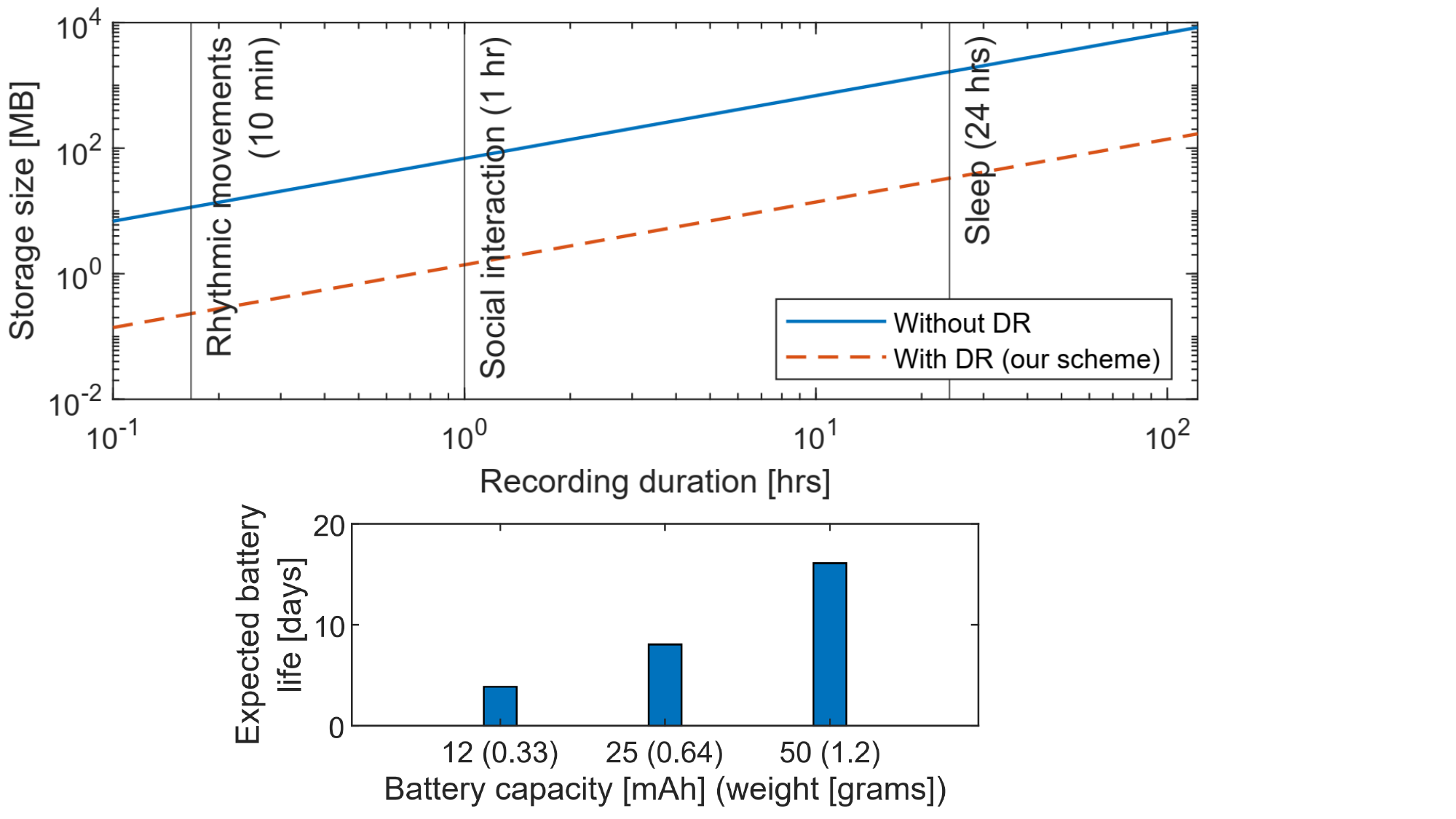}
    \centering
    \caption{Top: Storage-size savings when employing dimensionality reduction (DR) via the proposed approach for different types of cerebellar experiments. Bottom: Battery life of the head-stage when using the proposed approach for different miniature battery capacities \cite{powerstream2022}.}
    \label{fig:storage-savings}
\end{figure}

The results of the logic synthesis with the target clock frequency of 24.414 kHz (to match the sampling rate of the input neural data) are summarized in Table~\ref{tab:implementation-numbers}.
It can be seen that most of the energy is spent on the classifier during one complete detection-classification-storage cycle.
However, the classification is only invoked roughly a 100 times per second (Purkinje-cell firing rate), which results in a very-low system energy consumption and sufficiently long battery life (as will be discussed shortly).
In terms of area, the complete detector and classifier design, and the STT-RAM occupy just under 26 mm$^2$, which can be easily accommodated in $\sim$200-mm$^2$ PCBs typically used in mouse head stages~\cite{de2020ninscope}. Figure~\ref{fig:storage-savings} (top) shows that the dimensionality reduction performed by our classification scheme significantly reduces the storage requirements.

The next step is to calculate the battery life of the complete head stage in order to prove that it can stay operational throughout the longest of experiments (i.e., 24 hours).
Figure~\ref{fig:storage-savings} (bottom) shows the expected battery life of the head stage for three very-small-sized batteries~\cite{powerstream2022} using an ADC with 0.5 pJ per conversion~\cite{singh2021srif}. It can be seen that our approach easily allows for roughly 4 days of continuous operation for the smallest 0.33-gram battery (12 mAh).

\begin{table*}[!t]
    \centering
    \caption{Comparison with the state of the art}
    \label{tab:comparison}
    \begin{threeparttable}
    \begin{tabular}{l|l|l|l|l|l|l|l|l|l|l|l}
        \hline
          & \cite{freelynx2021manual} & \cite{bilodeau2021wireless} & \cite{cereplex2020manual} & \cite{ecube2020} & \cite{gagnon2016wireless} & \cite{grand2013long} & \cite{luan2018compact} &\cite{sedaghat2021p} & \cite{markanday2020using} & \cite{zur2019using}  & This work \\ 
        \hline
        CS-classification accuracy (\%) & \multicolumn{7}{c|}{N/A$_1$} & $-$ & $-$ & $-$ & 93.35 \\
        CS-classification F1 score (\%) & \multicolumn{7}{c|}{N/A$_1$} & $-$ & 92.5 & $-$ & 90.15 \\
        SS-classification accuracy (\%) & \multicolumn{7}{c|}{N/A$_1$} & $-$ & $-$ & $-$ & 96.67 \\
        SS-classification F1 score (\%) & \multicolumn{7}{c|}{N/A$_1$} & $-$ & $-$ & $-$ & 94.93 \\
        Commercial product & yes & no & yes & yes & no & no & no & no & no & no & no \\
        Purkinje-cell spike classification & no & no & no & no & no & no & no & yes & yes & yes & yes\\
        On-board DR & no & yes & no & no & yes & no & yes & \multicolumn{3}{c|}{N/A$_2$} & yes \\
        Weight (g) & $>$21 & 4.68 & 9.87 & 2.5 & 4.9 & $-$ & $-$ & \multicolumn{3}{c|}{N/A$_2$} & $<$2.5$^*$ \\
        Autonomy (hrs) & 3 & 1.6 & 2.5 & 0.5 & 1.75 & 72 & 24 & \multicolumn{3}{c|}{N/A$_2$} & 24$^{**}$\\
        Targeted animal & rats & mice & rats & mice & mice & cats & monkeys & \multicolumn{3}{c|}{N/A$_2$} & mice\\
        \hline
	\end{tabular}
	\begin{tablenotes}[flushleft]
		\item `$-$': Not provided, `N/A$_1$': Not applicable since these works do not deal with Purkinje cells, `N/A$_2$': Not applicable since these works are software based (i.e., not suitable for an animal head stage)
		\item $^{*}$Worst-case approximation based on the battery choice and design size.
		\item $^{**}$For the 24-hour-experiment configuration (32-MB STT-RAM).
	\end{tablenotes}
    \end{threeparttable}
\end{table*}



\subsection{Discussion}

Recall from Section~\ref{sec:introduction} that the goals of our scheme were to achieve (1) \textit{long-duration} experiments, and (2) \textit{freely moving} mice. The first goal can be validated from Figure~\ref{fig:storage-savings} in which we demonstrated that our scheme easily allows the head stage to operate continuously for the longest-duration experiments. Regarding the second goal, we saw that the use of an on-board storage allows us to get rid of the wires, which aids in the free movement of mice. Moreover, the design supports the use of very light batteries (down to 0.33 grams). Since the battery is the heaviest component of the head stage, our approach easily meets the requirement of the head-stage weight being less than 3 grams.
Table~\ref{tab:comparison} compares our work with the closely-related state of the art.
It can be seen that our scheme, to the best of our knowledge, is the only one to date that fulfills the aforementioned goals.
Moreover, it is the only one that performs small-form-factor Purkinje-cell spike classification.

\section{Conclusions}
\label{sec:conclusions}

In this paper, we proposed a lightweight architecture for classifying Purkinje-cell spikes from a stream of neural data.
Our scheme reduces the dimensionality of this very sparse data by classifying spikes with an overall accuracy of $>$95\%, which allows it to be stored on a removable storage on a mouse's head stage. This results in getting rid of the wires for data acquisition and enabling the free movement of mice during cerebellar-recording experiments. Moreover, our CMOS synthesis results demonstrate that our small-form-factor approach allows for long-duration experiments. The head stage utilizing our scheme can continuously run for approximately 4 days on a small 0.33-gram (12 mAh) battery. 
Such experiments involving free-moving mice (i.e., more realistic and natural conditions) can significantly help in elucidating the underlying mechanisms behind motor control and loss thereof, in the brain.


\begin{acks}
This paper is supported by the European Union’s Horizon Europe research and innovation programme under projects SEPTON (Gr. Agr. No. 101094901) and SECURED (Gr. Agr. No. 101095717) and by the Dutch Research Council's Gravitation programme under project DBI\textsuperscript{2} (No. 024.005.022).
\end{acks}

\bibliographystyle{ACM-Reference-Format}
\bibliography{References}


\begin{thebibliography}{31}


\ifx \showCODEN    \undefined \def \showCODEN     #1{\unskip}     \fi
\ifx \showDOI      \undefined \def \showDOI       #1{#1}\fi
\ifx \showISBNx    \undefined \def \showISBNx     #1{\unskip}     \fi
\ifx \showISBNxiii \undefined \def \showISBNxiii  #1{\unskip}     \fi
\ifx \showISSN     \undefined \def \showISSN      #1{\unskip}     \fi
\ifx \showLCCN     \undefined \def \showLCCN      #1{\unskip}     \fi
\ifx \shownote     \undefined \def \shownote      #1{#1}          \fi
\ifx \showarticletitle \undefined \def \showarticletitle #1{#1}   \fi
\ifx \showURL      \undefined \def \showURL       {\relax}        \fi
\providecommand\bibfield[2]{#2}
\providecommand\bibinfo[2]{#2}
\providecommand\natexlab[1]{#1}
\providecommand\showeprint[2][]{arXiv:#2}

\bibitem[Bertolazzi et~al\mbox{.}(2019)]%
        {bertolazzi2019nonvolatile}
\bibfield{author}{\bibinfo{person}{Simone Bertolazzi}, \bibinfo{person}{Paolo
  Bondavalli}, \bibinfo{person}{Stephan Roche}, \bibinfo{person}{Tamer San},
  \bibinfo{person}{Sung-Yool Choi}, \bibinfo{person}{Luigi Colombo},
  \bibinfo{person}{Francesco Bonaccorso}, {and} \bibinfo{person}{Paolo
  Samori}.} \bibinfo{year}{2019}\natexlab{}.
\newblock \showarticletitle{Nonvolatile memories based on graphene and related
  2D materials}.
\newblock \bibinfo{journal}{\emph{Advanced materials}} \bibinfo{volume}{31},
  \bibinfo{number}{10} (\bibinfo{year}{2019}), \bibinfo{pages}{1806663}.
\newblock


\bibitem[Bilodeau et~al\mbox{.}(2021)]%
        {bilodeau2021wireless}
\bibfield{author}{\bibinfo{person}{Guillaume Bilodeau},
  \bibinfo{person}{Gabriel Gagnon-Turcotte}, \bibinfo{person}{L{\'e}onard~L
  Gagnon}, \bibinfo{person}{Iason Keramidis}, \bibinfo{person}{Igor Timofeev},
  \bibinfo{person}{Yves De~Koninck}, \bibinfo{person}{Christian Ethier}, {and}
  \bibinfo{person}{Benoit Gosselin}.} \bibinfo{year}{2021}\natexlab{}.
\newblock \showarticletitle{A wireless electro-optic platform for multimodal
  electrophysiology and optogenetics in freely moving rodents}.
\newblock \bibinfo{journal}{\emph{Frontiers in Neuroscience}}
  \bibinfo{volume}{15} (\bibinfo{year}{2021}), \bibinfo{pages}{718478}.
\newblock


\bibitem[{Blackrock Microsystems, LLC}(2020)]%
        {cereplex2020manual}
\bibfield{author}{\bibinfo{person}{{Blackrock Microsystems, LLC}}.}
  \bibinfo{year}{2020}\natexlab{}.
\newblock \bibinfo{booktitle}{\emph{{CerePlex Exilis - Instructions for Use}}}.
\newblock


\bibitem[{Cambridge NeuroTech}(2020)]%
        {cambridge2020manual}
\bibfield{author}{\bibinfo{person}{{Cambridge NeuroTech}}.}
  \bibinfo{year}{2020}\natexlab{}.
\newblock \bibinfo{booktitle}{\emph{{Mini-Amp-64 User Guide -- Version 1.0}}}.
\newblock


\bibitem[Chung et~al\mbox{.}(2019)]%
        {chung2019high}
\bibfield{author}{\bibinfo{person}{Jason~E Chung}, \bibinfo{person}{Hannah~R
  Joo}, \bibinfo{person}{Jiang~Lan Fan}, \bibinfo{person}{Daniel~F Liu},
  \bibinfo{person}{Alex~H Barnett}, \bibinfo{person}{Supin Chen},
  \bibinfo{person}{Charlotte Geaghan-Breiner}, \bibinfo{person}{Mattias~P
  Karlsson}, \bibinfo{person}{Magnus Karlsson}, \bibinfo{person}{Kye~Y Lee},
  {et~al\mbox{.}}} \bibinfo{year}{2019}\natexlab{}.
\newblock \showarticletitle{High-density, long-lasting, and multi-region
  electrophysiological recordings using polymer electrode arrays}.
\newblock \bibinfo{journal}{\emph{Neuron}} \bibinfo{volume}{101},
  \bibinfo{number}{1} (\bibinfo{year}{2019}), \bibinfo{pages}{21--31}.
\newblock


\bibitem[de~Groot et~al\mbox{.}(2020)]%
        {de2020ninscope}
\bibfield{author}{\bibinfo{person}{Andres de Groot},
  \bibinfo{person}{Bastijn~JG van~den Boom}, \bibinfo{person}{Romano~M van
  Genderen}, \bibinfo{person}{Joris Coppens}, \bibinfo{person}{John van
  Veldhuijzen}, \bibinfo{person}{Joop Bos}, \bibinfo{person}{Hugo Hoedemaker},
  \bibinfo{person}{Mario Negrello}, \bibinfo{person}{Ingo Willuhn},
  \bibinfo{person}{Chris~I De~Zeeuw}, {et~al\mbox{.}}}
  \bibinfo{year}{2020}\natexlab{}.
\newblock \showarticletitle{NINscope, a versatile miniscope for multi-region
  circuit investigations}.
\newblock \bibinfo{journal}{\emph{Elife}}  \bibinfo{volume}{9}
  (\bibinfo{year}{2020}), \bibinfo{pages}{e49987}.
\newblock


\bibitem[De~Zeeuw et~al\mbox{.}(2011)]%
        {de2011spatiotemporal}
\bibfield{author}{\bibinfo{person}{Chris~I De~Zeeuw}, \bibinfo{person}{Freek~E
  Hoebeek}, \bibinfo{person}{Laurens~WJ Bosman}, \bibinfo{person}{Martijn
  Schonewille}, \bibinfo{person}{Laurens Witter}, {and}
  \bibinfo{person}{Sebastiaan~K Koekkoek}.} \bibinfo{year}{2011}\natexlab{}.
\newblock \showarticletitle{Spatiotemporal firing patterns in the cerebellum}.
\newblock \bibinfo{journal}{\emph{Nature Reviews Neuroscience}}
  \bibinfo{volume}{12}, \bibinfo{number}{6} (\bibinfo{year}{2011}),
  \bibinfo{pages}{327--344}.
\newblock


\bibitem[Dong et~al\mbox{.}(2012)]%
        {dong2012nvsim}
\bibfield{author}{\bibinfo{person}{Xiangyu Dong}, \bibinfo{person}{Cong Xu},
  \bibinfo{person}{Yuan Xie}, {and} \bibinfo{person}{Norman~P Jouppi}.}
  \bibinfo{year}{2012}\natexlab{}.
\newblock \showarticletitle{{NVSim: A circuit-level performance, energy, and
  area model for emerging nonvolatile memory}}.
\newblock \bibinfo{journal}{\emph{IEEE Transactions on Computer-Aided Design of
  Integrated Circuits and Systems}} \bibinfo{volume}{31}, \bibinfo{number}{7}
  (\bibinfo{year}{2012}), \bibinfo{pages}{994--1007}.
\newblock


\bibitem[{Everspin Technologies}(2024)]%
        {Everspin}
\bibfield{author}{\bibinfo{person}{{Everspin Technologies}}.}
  \bibinfo{year}{2024}\natexlab{}.
\newblock \bibinfo{booktitle}{\emph{{Spin-transfer Torque MRAM Technology}}}.
\newblock
\urldef\tempurl%
\url{https://www.everspin.com/}
\showURL{%
\tempurl}


\bibitem[Gagnon-Turcotte et~al\mbox{.}(2017)]%
        {gagnon2016wireless}
\bibfield{author}{\bibinfo{person}{Gabriel Gagnon-Turcotte},
  \bibinfo{person}{Yoan LeChasseur}, \bibinfo{person}{Cyril Bories},
  \bibinfo{person}{Youn{\`e}s Messaddeq}, \bibinfo{person}{Yves De~Koninck},
  {and} \bibinfo{person}{Benoit Gosselin}.} \bibinfo{year}{2017}\natexlab{}.
\newblock \showarticletitle{A wireless headstage for combined optogenetics and
  multichannel electrophysiological recording}.
\newblock \bibinfo{journal}{\emph{IEEE transactions on biomedical circuits and
  systems}} \bibinfo{volume}{11}, \bibinfo{number}{1} (\bibinfo{year}{2017}),
  \bibinfo{pages}{1--14}.
\newblock


\bibitem[Grand et~al\mbox{.}(2013)]%
        {grand2013long}
\bibfield{author}{\bibinfo{person}{Laszlo Grand}, \bibinfo{person}{Sergiu
  Ftomov}, {and} \bibinfo{person}{Igor Timofeev}.}
  \bibinfo{year}{2013}\natexlab{}.
\newblock \showarticletitle{Long-term synchronized electrophysiological and
  behavioral wireless monitoring of freely moving animals}.
\newblock \bibinfo{journal}{\emph{Journal of neuroscience methods}}
  \bibinfo{volume}{212}, \bibinfo{number}{2} (\bibinfo{year}{2013}),
  \bibinfo{pages}{237--241}.
\newblock


\bibitem[Jacob et~al\mbox{.}(2018)]%
        {jacob2018quantization}
\bibfield{author}{\bibinfo{person}{Benoit Jacob}, \bibinfo{person}{Skirmantas
  Kligys}, \bibinfo{person}{Bo Chen}, \bibinfo{person}{Menglong Zhu},
  \bibinfo{person}{Matthew Tang}, \bibinfo{person}{Andrew Howard},
  \bibinfo{person}{Hartwig Adam}, {and} \bibinfo{person}{Dmitry Kalenichenko}.}
  \bibinfo{year}{2018}\natexlab{}.
\newblock \showarticletitle{Quantization and training of neural networks for
  efficient integer-arithmetic-only inference}. In
  \bibinfo{booktitle}{\emph{Proceedings of the IEEE conference on computer
  vision and pattern recognition}}. \bibinfo{publisher}{IEEE},
  \bibinfo{pages}{2704--2713}.
\newblock


\bibitem[Kingma and Ba(2014)]%
        {kingma2014adam}
\bibfield{author}{\bibinfo{person}{Diederik~P Kingma} {and}
  \bibinfo{person}{Jimmy Ba}.} \bibinfo{year}{2014}\natexlab{}.
\newblock \showarticletitle{Adam: A method for stochastic optimization}.
\newblock \bibinfo{journal}{\emph{arXiv preprint arXiv:1412.6980}}
  (\bibinfo{year}{2014}).
\newblock


\bibitem[Luan et~al\mbox{.}(2018)]%
        {luan2018compact}
\bibfield{author}{\bibinfo{person}{Song Luan}, \bibinfo{person}{Ian Williams},
  \bibinfo{person}{Michal Maslik}, \bibinfo{person}{Yan Liu},
  \bibinfo{person}{Felipe De~Carvalho}, \bibinfo{person}{Andrew Jackson},
  \bibinfo{person}{Rodrigo~Quian Quiroga}, {and} \bibinfo{person}{Timothy~G
  Constandinou}.} \bibinfo{year}{2018}\natexlab{}.
\newblock \showarticletitle{Compact standalone platform for neural recording
  with real-time spike sorting and data logging}.
\newblock \bibinfo{journal}{\emph{Journal of neural engineering}}
  \bibinfo{volume}{15}, \bibinfo{number}{4} (\bibinfo{year}{2018}),
  \bibinfo{pages}{046014}.
\newblock


\bibitem[Markanday et~al\mbox{.}(2020)]%
        {markanday2020using}
\bibfield{author}{\bibinfo{person}{Akshay Markanday}, \bibinfo{person}{Joachim
  Bellet}, \bibinfo{person}{Marie~E Bellet}, \bibinfo{person}{Junya Inoue},
  \bibinfo{person}{Ziad~M Hafed}, {and} \bibinfo{person}{Peter Thier}.}
  \bibinfo{year}{2020}\natexlab{}.
\newblock \showarticletitle{Using deep neural networks to detect complex spikes
  of cerebellar Purkinje cells}.
\newblock \bibinfo{journal}{\emph{Journal of neurophysiology}}
  \bibinfo{volume}{123}, \bibinfo{number}{6} (\bibinfo{year}{2020}),
  \bibinfo{pages}{2217--2234}.
\newblock


\bibitem[McInnes et~al\mbox{.}(2018)]%
        {mcinnes2018umap}
\bibfield{author}{\bibinfo{person}{Leland McInnes}, \bibinfo{person}{John
  Healy}, {and} \bibinfo{person}{James Melville}.}
  \bibinfo{year}{2018}\natexlab{}.
\newblock \showarticletitle{{UMAP: Uniform manifold approximation and
  projection for dimension reduction}}.
\newblock \bibinfo{journal}{\emph{arXiv preprint arXiv:1802.03426}}
  (\bibinfo{year}{2018}).
\newblock


\bibitem[Mukhopadhyay and Ray(1998)]%
        {mukhopadhyay1998new}
\bibfield{author}{\bibinfo{person}{Sudipta Mukhopadhyay} {and}
  \bibinfo{person}{GC Ray}.} \bibinfo{year}{1998}\natexlab{}.
\newblock \showarticletitle{A new interpretation of nonlinear energy operator
  and its efficacy in spike detection}.
\newblock \bibinfo{journal}{\emph{IEEE Transactions on biomedical engineering}}
  \bibinfo{volume}{45}, \bibinfo{number}{2} (\bibinfo{year}{1998}),
  \bibinfo{pages}{180--187}.
\newblock


\bibitem[Nair and othes(2018)]%
        {nair2018defect}
\bibfield{author}{\bibinfo{person}{Sarath~Mohanachandran Nair} {and}
  \bibinfo{person}{othes}.} \bibinfo{year}{2018}\natexlab{}.
\newblock \showarticletitle{{Defect injection, fault modeling and test
  algorithm generation methodology for STT-MRAM}}. In
  \bibinfo{booktitle}{\emph{International Test Conference (ITC)}}. IEEE,
  \bibinfo{pages}{1--10}.
\newblock


\bibitem[{NeuraLynx, Inc.}(2021)]%
        {freelynx2021manual}
\bibfield{author}{\bibinfo{person}{{NeuraLynx, Inc.}}}
  \bibinfo{year}{2021}\natexlab{}.
\newblock \bibinfo{booktitle}{\emph{{FreeLynx - User Manual}}}.
\newblock


\bibitem[Oboril et~al\mbox{.}(2015)]%
        {oboril2015evaluaion}
\bibfield{author}{\bibinfo{person}{Fabian Oboril}, \bibinfo{person}{Rajendra
  Bishnoi}, \bibinfo{person}{Mojtaba Ebrahimi}, {and} \bibinfo{person}{Mehdi~B
  Tahoori}.} \bibinfo{year}{2015}\natexlab{}.
\newblock \showarticletitle{Evaluation of hybrid memory technologies using
  SOT-MRAM for on-chip cache hierarchy}.
\newblock \bibinfo{journal}{\emph{IEEE Transactions on Computer-Aided Design of
  Integrated Circuits and Systems}} \bibinfo{volume}{34}, \bibinfo{number}{3}
  (\bibinfo{year}{2015}), \bibinfo{pages}{367--380}.
\newblock


\bibitem[{PowerStream Technology}(2022)]%
        {powerstream2022}
\bibfield{author}{\bibinfo{person}{{PowerStream Technology}}.}
  \bibinfo{year}{2022}\natexlab{}.
\newblock \bibinfo{booktitle}{\emph{{Ultra low weight lithium ion batteries}}}.
\newblock
\urldef\tempurl%
\url{https://www.powerstream.com/ultra-light.htm}
\showURL{%
\tempurl}


\bibitem[Roeder(2022)]%
        {netron2022}
\bibfield{author}{\bibinfo{person}{Lutz Roeder}.}
  \bibinfo{year}{2022}\natexlab{}.
\newblock \bibinfo{booktitle}{\emph{{Netron}}}.
\newblock
\urldef\tempurl%
\url{https://github.com/lutzroeder/netron}
\showURL{%
\tempurl}


\bibitem[Romano et~al\mbox{.}(2020)]%
        {romano2020functional}
\bibfield{author}{\bibinfo{person}{Vincenzo Romano},
  \bibinfo{person}{Aoibhinn~L Reddington}, \bibinfo{person}{Silvia Cazzanelli},
  \bibinfo{person}{Roberta Mazza}, \bibinfo{person}{Yang Ma},
  \bibinfo{person}{Christos Strydis}, \bibinfo{person}{Mario Negrello},
  \bibinfo{person}{Laurens~WJ Bosman}, {and} \bibinfo{person}{Chris~I
  De~Zeeuw}.} \bibinfo{year}{2020}\natexlab{}.
\newblock \showarticletitle{Functional convergence of autonomic and
  sensorimotor processing in the lateral cerebellum}.
\newblock \bibinfo{journal}{\emph{Cell reports}} \bibinfo{volume}{32},
  \bibinfo{number}{1} (\bibinfo{year}{2020}), \bibinfo{pages}{107867}.
\newblock


\bibitem[Salahuddin et~al\mbox{.}(2018)]%
        {salahuddin2018era}
\bibfield{author}{\bibinfo{person}{Sayeef Salahuddin} {et~al\mbox{.}}}
  \bibinfo{year}{2018}\natexlab{}.
\newblock \showarticletitle{The era of hyper-scaling in electronics}.
\newblock \bibinfo{journal}{\emph{Nature Electronics}} \bibinfo{volume}{1},
  \bibinfo{number}{8} (\bibinfo{year}{2018}), \bibinfo{pages}{442--450}.
\newblock


\bibitem[Sedaghat-Nejad et~al\mbox{.}(2021)]%
        {sedaghat2021p}
\bibfield{author}{\bibinfo{person}{Ehsan Sedaghat-Nejad},
  \bibinfo{person}{Mohammad~Amin Fakharian}, \bibinfo{person}{Jay Pi},
  \bibinfo{person}{Paul Hage}, \bibinfo{person}{Yoshiko Kojima},
  \bibinfo{person}{Robi Soetedjo}, \bibinfo{person}{Shogo Ohmae},
  \bibinfo{person}{Javier~F Medina}, {and} \bibinfo{person}{Reza Shadmehr}.}
  \bibinfo{year}{2021}\natexlab{}.
\newblock \showarticletitle{P-sort: an open-source software for cerebellar
  neurophysiology}.
\newblock \bibinfo{journal}{\emph{Journal of neurophysiology}}
  \bibinfo{volume}{126}, \bibinfo{number}{4} (\bibinfo{year}{2021}),
  \bibinfo{pages}{1055--1075}.
\newblock


\bibitem[{Silicon Integration Initiative, Inc}(2019)]%
        {nangate2019}
\bibfield{author}{\bibinfo{person}{{Silicon Integration Initiative, Inc}}.}
  \bibinfo{year}{2019}\natexlab{}.
\newblock \bibinfo{booktitle}{\emph{{15nm Open-Cell Library and 45nm
  FreePDK}}}.
\newblock
\urldef\tempurl%
\url{https://si2.org/open-cell-library/}
\showURL{%
\tempurl}


\bibitem[Singh et~al\mbox{.}(2021)]%
        {singh2021srif}
\bibfield{author}{\bibinfo{person}{Abhairaj Singh}, \bibinfo{person}{Muath~Abu
  Lebdeh}, \bibinfo{person}{Anteneh Gebregiorgis}, \bibinfo{person}{Rajendra
  Bishnoi}, \bibinfo{person}{Rajiv~V Joshi}, {and} \bibinfo{person}{Said
  Hamdioui}.} \bibinfo{year}{2021}\natexlab{}.
\newblock \showarticletitle{{SRIF: Scalable and reliable integrate and fire
  circuit adc for memristor-based cim architectures}}.
\newblock \bibinfo{journal}{\emph{IEEE Transactions on Circuits and Systems I:
  Regular Papers}} \bibinfo{volume}{68}, \bibinfo{number}{5}
  (\bibinfo{year}{2021}), \bibinfo{pages}{1917--1930}.
\newblock


\bibitem[Smullen et~al\mbox{.}(2011)]%
        {smullen2011relaxing}
\bibfield{author}{\bibinfo{person}{Clinton~W Smullen},
  \bibinfo{person}{Vidyabhushan Mohan}, \bibinfo{person}{Anurag Nigam},
  \bibinfo{person}{Sudhanva Gurumurthi}, {and} \bibinfo{person}{Mircea~R
  Stan}.} \bibinfo{year}{2011}\natexlab{}.
\newblock \showarticletitle{Relaxing non-volatility for fast and
  energy-efficient STT-RAM caches}. In \bibinfo{booktitle}{\emph{2011 IEEE 17th
  International Symposium on High Performance Computer Architecture}}. IEEE,
  \bibinfo{pages}{50--61}.
\newblock


\bibitem[{White Matter, LLC}(2020)]%
        {ecube2020}
\bibfield{author}{\bibinfo{person}{{White Matter, LLC}}.}
  \bibinfo{year}{2020}\natexlab{}.
\newblock \bibinfo{booktitle}{\emph{{Details of the eCube standalone and
  stacking headstages}}}.
\newblock
\urldef\tempurl%
\url{https://docs.white-matter.com/docs/ecube/hardware/headstages/}
\showURL{%
\tempurl}


\bibitem[Yang and Mason(2016)]%
        {yang2016hardware}
\bibfield{author}{\bibinfo{person}{Yuning Yang} {and} \bibinfo{person}{Andrew~J
  Mason}.} \bibinfo{year}{2016}\natexlab{}.
\newblock \showarticletitle{Hardware efficient automatic thresholding for
  NEO-based neural spike detection}.
\newblock \bibinfo{journal}{\emph{IEEE Transactions on Biomedical Engineering}}
  \bibinfo{volume}{64}, \bibinfo{number}{4} (\bibinfo{year}{2016}),
  \bibinfo{pages}{826--833}.
\newblock


\bibitem[Zur and Joshua(2019)]%
        {zur2019using}
\bibfield{author}{\bibinfo{person}{Gil Zur} {and} \bibinfo{person}{Mati
  Joshua}.} \bibinfo{year}{2019}\natexlab{}.
\newblock \showarticletitle{Using extracellular low frequency signals to
  improve the spike sorting of cerebellar complex spikes}.
\newblock \bibinfo{journal}{\emph{Journal of Neuroscience Methods}}
  \bibinfo{volume}{328} (\bibinfo{year}{2019}), \bibinfo{pages}{108423}.
\newblock


\end{thebibliography}

\end{document}